\documentclass{elsart}

\usepackage{amssymb}
\usepackage{graphicx}

\begin{document}

\voffset = 0.0cm

\begin{frontmatter}

%%----------------------------------------------------------------------------------
\title{\rm Neutron-rich rare isotope production with stable and radioactive beams 
in the mass range A$\sim$40--60 at beam energy around 15 MeV/nucleon}

%%----------------------------------------------------------------------------------
%%%
%%% Author list in elsart style:
%%%

\author{ A. Papageorgiou$^{1}$, G.A. Souliotis$^{1,*}$, K. Tshoo$^{2}$, S.C. Jeong$^{2}$, B.H. Kang$^{2}$,}

\author{Y.K. Kwon$^{2}$, M. Veselsky$^{3}$, S.J. Yennello$^{4}$, and A. Bonasera$^{4,5}$ }

%%-----------------------------------------------------------------------------------

\address{ $^{1}$ Laboratory of Physical Chemistry, Department of Chemistry,
                 National and Kapodistrian University of Athens, Athens 15771, Greece }
    
\address{ $^{2}$ The Rare Isotope Science Project (RISP), Institute for Basic Science, Daejeon, Korea.}
                             
\address{ $^{3}$ Institute of Physics, Slovak Academy of
                 Sciences, Bratislava 84511, Slovakia }
           
\address{ $^{4}$ Cyclotron Institute, Texas A\&M University, College Station, Texas, USA}              

\address{ $^{5}$ Laboratori Nazionali del Sud, INFN, Catania, Italy }

\footnote{ $^{*}$ Corresponding author. Email: soulioti@chem.uoa.gr }

%%
%% \address{ $^{2}$ Department of Physics, University of Ioannina, Ioannina,  Greece }
%%
%%************************************************************************
%%************************************************************************
%abstract
%*************************************************************************

\begin{abstract}

%%%----------------
%% We reports on our continued efforts to systematically study the production 
%% of rare isotopes with heavy-ion beams at energy of 15 MeV/nucleon. 
%%%----------------
We studied the production of neutron-rich nuclides in multinucleon
transfer collisions of stable and radioactive beams in the mass range A$\sim$40--60. 
%%%
We first presented our experimental cross section data of projectile fragments from 
the reaction of $^{40}$Ar(15 MeV/nucleon) with $^{64}$Ni, $^{58}$Ni and $^{27}$Al.
%%%
We then compared them with calculations based on either  the deep-inelastic transfer (DIT)
model or the constrained molecular  dynamics (CoMD) model, followed by  the statistical 
multifragmentation model (SMM).  An overall good agreement of the calculations with the 
experimental data is obtained.  
%%%
We continued with calculations of the reaction of $^{40}$Ar (15 MeV/nucleon) 
with $^{238}$U target and then with reactions of $^{48}$Ca (15 MeV/nucleon)
with $^{64}$Ni and $^{238}$U targets. 
%%%
In these reactions,  neutron-rich rare isotopes with large cross sections are produced.
%%%
These nuclides, in turn, can be assumed to form radioactive beams and interact with
a subsequent target  (preferably $^{238}$U), leading to the production of extremely
neutron-rich and even new isotopes  (e.g. $^{60}$Ca) in this  mass range.   %%%% of A=40--60.
%%%
We conclude that multinucleon transfer reactions with stable or radioactive beams 
at the energy of around 15 MeV/nucleon 
offer an effective route to access extremely neutron-rich rare isotopes  for nuclear 
structure or reaction studies.
%%%
\end{abstract}

\end{frontmatter}

%%*************************************************************************************************
% main text

\section{Introduction}

%%%***********************************************************************************************
%%%
The study of the chart of the nuclides toward the  astrophysical
r-process path and the neutron drip-line continues to receive special 
attention by the nuclear physics community  
(see, e.g., \cite{Utama-2017,Afanasjev-2015,Gade-2015,ndrip0}). 
%%%
Moreover,  the efficient production of  very neutron-rich nuclides constitutes 
a central issue in current and upcoming rare isotope beam facilities
(see, e.g., \cite{FRIB1,FRIB2,GANIL,GSI,RIBF,ARGONNE,EURISOL,RISP,RISP-2013}).

Neutron-rich nuclides  are routinely produced  by spallation, fission or projectile  
fragmentation \cite{RIB-2013}.
%%%
Spallation is a traditional mechanism to produce rare isotopes for ISOL-type 
techniques \cite{Spallation}.
%%%
Projectile fission has proven very effective to produce light and
heavy fission fragments (see, e.g., \cite{Ufission,Vonta-2016,Flavigny-2017}).
%%%
Finally, projectile fragmentation has been established as a universal approach to produce 
exotic nuclei at beam energies typically above 100 MeV/nucleon 
(see, e.g., \cite{MSUfrag1,MSUfrag2,Kurtukian-2014,Caballero-2016,Samantaria-2015,Meisel-2016}).
%%%
We point out that in this approach, optimum neutron excess in the fragments is achieved 
by stripping the maximum possible number of protons (and a minimum number of neutrons).

To arrive at an even  higher neutron-excess in the products, apart from proton stripping,
capture of neutrons from the target is necessary.
Such a possibility is offered by  reactions involving nucleon exchange  at beam energies
from the Coulomb barrier \cite{Volkov,Corradi,Corradi-2013} to the Fermi energy
(below 40 MeV/nucleon) \cite{GS-PLB-2002,GS-PRL-2003}.
%%%
There is renewed interest in such reactions nowadays as evidenced by recent experimental
(e.g. \cite{Beliuskina-2014,Watanabe-2015,Vogt-2015,Mijatovic-2016,Welsh-2017}) 
and theoretical (e.g. \cite{Zhu-2017,Sekizawa-2017a,Sekizawa-2017b,Karpov-2017,Wang-2016,Wang-2017})
works. 
%%%
These efforts focus mainly on multinucleon transfer reactions near the Coulomb barrier.
From a practical standpoint,
in such reactions, the low velocities of the fragments  and the wide angular and ionic 
charge state distributions may limit the collection efficiency for the most neutron-rich products.

%%%---------------------------------------------------------
The reactions in the Fermi energy regime  (i.e. 15--35 MeV/nucleon) %%% \cite{Fuchs}
combine  the advantages  of both low-energy (i.e., near and above the Coulomb barrier) 
and high-energy (i.e., above 100 MeV/nucleon) reactions.
%%%
At this energy, the overlap of the peripheries of the projectile and the target
enhances the N/Z of the resulting fragments, while the velocities are  high enough to
allow efficient in-flight collection and separation.

Our previous experimental studies of projectile fragments from 15 and 25 MeV/nucleon 
reactions  of $^{86}$Kr \cite{GS-PLB-2002,GS-PRL-2003,GS-NIM-2003,GS-NIM-2008,GS-PRC-2011}
indicated substantial  production of neutron-rich fragments.
%%%
Our recent article \cite{Fountas-2014}  elaborates on our current understanding of the reaction
mechanism and our ability to describe it quantitatively with the  phenomenological DIT (deep
inelastic transfer)  model, as well as with the microscopic CoMD (Constrained Molecular Dynamics)
model.
%%% QP-fragmentation:
%%% For the purpose of this work, we may call this mechanism quasi-projectile (QP) fragmentation to 
%%% make a clear distinction from the widely applied high-energy projectile fragmentation method.
%%%
As already pointed out in our previous works, our multinucleon transfer approach in the energy regime 
of 15--25 MeV/nucleon offers the  possibility of 
essentially  adding neutrons (along  with the usual stripping of protons) to a given stable
(or radioactive) projectile via interaction with a neutron-rich target. 
%%%

In this article, after a short overview of our previous experimental measurements
with an $^{40}$Ar (15 MeV/nucleon) beam, we present systematic calculations of the 
production cross sections based on either the phenomenological DIT model or 
the microscopic CoMD model.
%%% 
The good description of the experimental results with both the microscopic CoMD code, 
as well as, with the phenomenological DIT code,  suggest the possibility of using the
present theoretical framework for the prediction of exotic nuclei employing 
radioactive beams that  will soon be available  in upcoming facilities.  
%%%
As examples, we present the  production cross sections and the rates  
of neutron-rich nuclei using radioactive beams of  $^{46}$Ar and $^{54}$Ca at 15 MeV/nucleon.
%%%
The structure of the paper is as follows. In section 2, we give a short 
description of the experimental methods and in section 3  an overview of the models used.
In sections 4--6 we compare our calculations to data for the stable beams
$^{40}$Ar and $^{48}$Ca. Then, in sections 7 and 8 we present calculations with the
radioactive beams $^{46}$Ar and $^{54}$Ca. Finally,  we close  with a discussion and summary.
 
%%%************************************************************************************************
%%%	
%%%***********************************************************************************************
%%%%
\section{Brief description of the experimental data and aparatus}

Experimental data on neutron rich nuclide production with a beam of $^{40}$Ar (15 MeV/nucleon) 
were obtained at the  Cyclotron Institute of Texas A\&M University, in parallel to a series
of measurements with a $^{86}$Kr (15 MeV/nucleon) beam already published in \cite{GS-PRC-2011}.
%%%
A preliminary version of the $^{40}$Ar data has already been presented in \cite{TAMU-2009}.
%%%
The experimental setup has been presented in detail in \cite{GS-PRC-2011}.
%%%
For completeness, we give a brief overview of the experimental methods here. 
%%%
A 15 MeV/nucleon $^{40}$Ar$^{9+}$ beam hit  targets of  $^{64}$Ni, $^{58}$Ni and $^{27}$Al
with thickness of 2 mg/cm$^2$. Projectile fragments were collected and identified using 
the MARS recoil separator applying the techniques developed and documented in \cite{GS-PRC-2011}. 
%%%
The $^{40}$Ar beam was send on the primary target location of MARS with an inclination of 4$^o$
with respect to the optical axis of the separator and projectile fragments were collected in the
polar angular range of 2.2--5.5$^o$ (in a solid angle of $\Delta \Omega$ = 4.0 msr).
%%%
After interaction with the target, the fragments traversed a PPAC (parallel-plate avalanche 
counter) at the intermediate image location (for position and magnetic rigidity measurement 
and START-time information) and then they were focused at the end of the device passing through
a second PPAC (for image size monitoring and STOP-time information). 
%%%
Finally the fragments were collected in a 5x5 cm$^2$ $\Delta$E--E Si detector telescope 
(with 60 and 1000 $\mu$m thickness, respectively).
%%%
Following standard techniques of B$\rho$--$\Delta$E--E--TOF (magnetic rigidity, energy-loss, 
residual energy and time-of-flight, respectively), the atomic number Z, the mass number A, 
the velocity and the ionic charge  of the fragments were obtained on an event-by-event basis.
%%%
Data were obtained in a series of overlapping magnetic rigidity settings of the spectrometer  
in the range  1.1-1.5 Tm. 
%%%(GEORGE check) %%% to cover the energy and charge state distributions of the fragments.  
%%%
We note that this magnetic rigidity range did not fully cover the  neutron-deficient side
of the product distributions which extends down to approximately 0.8 Tm according to our
calculations with the models discussed in the following.
%%%
(The neutron-deficient isotopes with incomplete  B$\rho$ coverage lie to the left of the
thin solid lines in Figs. 1,2,3 and 6.)

%%%
In order to obtain total cross sections of the produced isotopes, we applied 
corrections to the measured yields for the limited angular coverage of the  
spectrometer, as performed in our previous work \cite{GS-PRC-2011}.
The corrections were based on simulations of the reactions using the DIT code (see below)
followed by a deexcitation code \cite{GS-PRC-2011}.
%%%
We used the ratio of the filtered to unfiltered calculated yields to correct the
measured yield data  (obtained in the limited angular acceptance of the spectrometer)  
and to extract the total production cross section for each isotope
that we present and discuss in the following.

%%%-----------------------------------------------------------------------------------------------
%%%Expt data 40Ar+Ni,Al
%%%
In figure 1 we present the extracted cross sections of the isotopes of elements Z=19--12
for the three reactions studied:  $^{40}$Ar (15 MeV/nucleon) + $^{64}$Ni, $^{58}$Ni and $^{27}$Al,
represented by closed (black) circles, open (red) circles and open (blue) squares,   
respectively. 
%%%
We clearly observe enhanced production of neutron-rich isotopes with the more 
neutron-rich target of $^{64}$Ni (N/Z=1.28),  followed by $^{58}$Ni (N/Z=1.07) 
and, finally, by $^{27}$Al (N/Z=1.08).
%%%
Interestingly, few-neutron pickup nuclides are produced in all three reactions.
These nuclides lie to the right of the vertical dotted lines in figure 1
(and subsequent figures). As expected, the reaction with the most neutron-rich 
of the three targets leads to the highest production of these nuclides in the data.
%%% 
Especially for elements very close to the projectile, e.g. Z=19--16, the neutron-pickup
possibility provides a distinct advantage of these reactions in comparison to the 
well-established projectile fragmentation approach, in which the neutron pickup
possibility is suppressed, as we will also discuss in the following. %%%% \cite{RIB-2013}.
 
%%%------------------------------------------------------------------------------------------------
\section{Outline of the theoretical models}

The calculations performed in this work are based on a two-stage Monte Carlo approach.
The dynamical stage of the collision  was described by two different models:
%%% DIT:
First, we employed the phenomenological deep-inelastic transfer (DIT) model \cite{DIT,DITm,Veselsky-2011} 
simulating stochastic nucleon exchange in peripheral collisions.
This model has been successful in describing the N/Z, excitation energy, and kinematical properties 
of excited quasiprojectiles in a number of recent studies (e.g. \cite{Souliotis-2014} and references therein), 
including our work on rare isotope production below the Fermi energy, as referenced in the 
previous section.

%%% COMD:
We also used the microscopic constrained molecular dynamics (CoMD) model 
\cite{CoMD1,CoMD2,Papa-2013,Guiliani-2014,Vonta-2015} 
successfully used in studies of low-energy reaction dynamics. 
This code follows the general approach of the quantum molecular dynamics (QMD) models \cite{QMD}
and describes the nucleons as localized Gaussian wave packets. 
It implements an effective nucleon-nucleon interaction with a nuclear-matter compressibility 
of K = 200 (soft EOS) and several forms of the density dependence of the nucleon-nucleon symmetry
potential. Moreover, it imposes a phase space constraint to restore the Pauli principle during
the time evolution of the system. 
%%%
As in our recent studies at 15 and 25 MeV/nucleon \cite{Fountas-2014,Souliotis-2010},
in this work we mainly employed a symmetry potential proportional to the density, 
(that we called the standard symmetry potential). We also performed calculations with a symmetry potential
proportional to the square root of the density (soft symmetry potential), as well as with one proportional 
to the square of the density (hard symmetry potential). Both of them lead to nearly similar results 
for the peripheral collisions involved in this study and are not shown in the paper.   
%%%
In the present CoMD calculations, the dynamical evolution of the system was 
stopped at t = 300 fm/c (10$^{-21}$ s) \cite{Fountas-2014}.

%%% Deexcitation:
After the dynamical stage simulated either by the DIT code or the CoMD code, the deexcitation stage
of the reaction was described by the  statistical code SMM (Statistical Multifragmentation Model)
of Botvina \cite{SMM1,SMM2}. In this code, thermally equilibrated partitions of hot fragments are 
generated in a hot stage, which is followed by the propagation of the fragments in their mutual Coulomb 
field  and their secondary deexcitation as they fly in their asymptotic directions. 
%%%
We note  that for low excitation energy events (E$^*$ $<$ 1 MeV/nucleon) of relevance to the 
production of very neutron-rich nuclei, the SMM code describes  adequately the deexcitation
process as a cascade of emissions of neutrons and light charged particles using the 
Weisskopf-Ewing model of statistical evaporation \cite{DITm,Veselsky-2011}.

%%% Comments on DIT and CoMD
%%%
After this brief outline the models, we wish to comment on the physical basis of the two models
used to describe the dynamical stage. The DIT model is a  phenomenological nucleon-exchange model 
with empirical parameters carefully chosen to describe peripheral reactions in the Fermi energy 
and below.
%%%
On the other hand, the CoMD model is a fully microscopic (semiclassical) N-body model employing 
empirical interactions among the nucleons adjusted to describe the known static properties of nuclei 
(i.e., radii, masses, etc.). As such, the code has essentially no adjustable parameters that depend
on the reaction dynamics. Moreover, the CoMD approach, contrary to mean-field models,  naturally 
takes into account correlations among nucleons that are important to describe  observables involving 
fluctuations, as for example the nucleon tranfer in peripheral heavy-ion collisions. 

We mention that multinucleon transfer in reactions near the Coulomb barrier has  been described with
some success by the fully quantal time-dependent Hartree-Fock (TDHF)  approach 
(see, e.g., Ref. \cite{Sekizawa-2017a,Sekizawa-2017b} and references therein), which is, however 
not appropriate for energies well above the Coulomb barrier. 
%%%
For the above reasons, we understand that the microscopic CoMD model offers a valuable 
theoretical microscopic framework for the description of the present reactions. 
Furthermore, the CoMD may be reliably applied to situations where no experimental
data are available, especially in reactions with neutron-rich rare isotope beams.

%%%
%%%------------------------------------------------------------------------------------------------

\section{Reactions with the $^{40}$Ar projectile at 15 MeV/nucleon}
%%%%%
%%%%% 40-Ar:
%%%%%
In this section, we first present  comparisons  of the calculated production cross
sections of $^{40}$Ar projectile fragments with our experimental data. 
%%%
Since our goal is toward the optimum production of neutron-rich isotopes, 
the comparisons will focus on the reaction of $^{40}$Ar with the $^{64}$Ni target.
Similar comparisons were performed with the other two  targets  (resulting in equivalent 
quality of agreement) and are not reported in this paper.

%%%We will then present calculations with the heavy target of $^{232}$U (N/Z=1.59).
%%%--------------------------------------------------------------------------------------------
%%% 40Ar+64Ni:  data, DIT/SMM, CoMD/SMM: (new) Fig. 2
%%%
In figure 2, we show the calculated mass distributions of projectile fragments with Z=19--12
from the reaction $^{40}$Ar (15 MeV/nucleon) + $^{64}$Ni obtained by DIT/SMM [solid (red) line] 
and by CoMD/SMM [dotted (blue) line] and compare them with the experimental data (closed points) 
described above (figure 1).
%%%
We observe that the results of the two models are almost identical and in good agreement with 
the experimental data, especially for isotopes close to the projectile (Z=19--16).
%%%
As mentioned in section 2,
the discrepancy between the calculation and the experimental data on the neutron-deficient side
is due to incomplete coverage of the magnetic rigidity in the experiment 
that resulted  in loss or incomplete collection of the neutron-deficient nuclides.
%%%
Moreover, the agreement of the DIT calculations with those with the microscopic CoMD model, 
as well as with the data  gives us confidence in the use of the DIT code for further 
systematic study of reactions for which no experimental data exist at present.

%%%--------------------------------------------------------------------------------------------
%%% 40Ar+64Ni, Pb, U:  data, DIT/SMM Fig. 3
%%%
In figure 3, we proceed to show the DIT/SMM calculations for the reaction of $^{40}$Ar (15 MeV/nucleon) 
with the heavy and very neutron-rich $^{238}$U (N/Z=1.59) target [dashed (blue) line] along with  
the DIT/SMM calculations [solid (red) line]  of $^{40}$Ar+$^{64}$Ni and the experimental data  (closed points)
that we already presented in figure 2.
%%%
As we might expect, the heavy $^{238}$U target leads to further enhanced production  of neutron-rich 
isotopes, especially at the tails of the distributions, as compared to those with the $^{64}$Ni target.
%%%
For further comparison in figure 3, we also show the projectile fragmentation data of Notani 
\cite{Notani-2000,Notani-2007} for the reaction $^{40}$Ar (94 MeV/nucleon) + $^{181}$Ta 
(open circles). 
%%%
In these data we observe the absence of neutron-pickup nuclides. However, such products 
are reported in the fragmentation data of Zhang et al. \cite{Zhang-2012} with a beam of 
$^{40}$Ar (57 MeV/nucleon) on $^{181}$Ta (shown in figure 3 by open diamonds) and $^{9}$Be targets,
as well as the data of Kwan et al. \cite{Kwan-2012} for the same reactions at 140 MeV/nucleon.
These pick-up products have cross sections lower by a factor of 10 or more compared to 
the present 15 MeV/nucleon reactions.
%%%
Qualitatively, this can be attributed  to the short contact time of the  projectile with the target 
that prevents extensive nucleon exchange,   which of course takes place  in the lower energy 
(e.g. 15 MeV/nucleon) reactions.
%%%	
From the above comparisons, we understand that the reactions of choice at our energy of 15 MeV/nucleon
are those with the neutron-rich target of $^{238}$U, since they lead  to especially enhanced
production of the most neutron-rich isotopes.

It is interesting to note that the reaction with the  $^{238}$U target leads to substantially enhanced 
cross sections for neutron-rich nuclides far from the projectile,  e.g. for Si, Al and Mg isotopes in figure 3.
The cross sections for the most neutron-rich of these isotopes (e.g.  $^{35}$Al,  $^{32}$Mg) are more than 
two orders of magnitute larger than those obtained with the $^{64}$Ni target in our reactions at 15 MeV/nucleon
(or with the projectile fragmentation reactions).
%%%
This enhanced production has been experimentally observed in our previous work with 
 $^{86}$Kr(25MeV/nucleon)+$^{208}$Pb  and successfully described with the DIT/SMM model framework
mainly as asymmetric binary decay of extremely neutron-rich quasiprojectiles
\cite{GS-PRC-2007,GS-NIM-2007}.

%%%*****************************************************************************************************
%%%*****************************************************************************************************

\section{Angular distributions of projectile fragments from $^{40}$Ar reactions at 15 MeV/nucleon}
%%%
%%% Ang. Dist 40Ar+ Ni, U  (DIT/SMM)   Fig. 4
%%%
To understand the kinematics and the angular spread of the fragments  from the reactions
of the $^{40}$Ar projectile  at 15 MeV/nucleon, in figure 4(a) we show the DIT/SMM calculated 
mass-resolved angular distributions for the $^{40}$Ar+$^{64}$Ni system for which the 
grazing angle is 7.0$^o$.
%%%
The successive contours (starting from the innermost) represent a decrease in the yield by a factor of two. 
The pair of  horizontal full  lines represent the polar angular acceptance  of the MARS spectrometer
in the experimental setup of ref. \cite{GS-PRC-2011} (section 2). The pair of horizontal dashed lines
indicate the angular	acceptance of the KOBRA separator \cite{KOBRA-2016}, that we consider as a 
representative large-acceptance separator especially designed for rare-isotope production at this energy 
regime in the upcoming RIB facility RISP \cite{RISP}.
%%%
In the KOBRA setup, we assume that the beam hits the primary target at an angle of 5$^{o}$  (that may be
achieved with the aid of an appropriate beam swinger system) and fragments are collected in the polar 
angular range of 0$^{o}$--10$^{o}$ in a solid angle of $\Delta \Omega$ = 50 msr.

From the figure, we qualitatively appreciate the issue of the relatively limited angular acceptance
of a medium acceptance separator like  MARS. Our simulations indicate that a fraction of $\sim$5\%  %%%or smaller
(taking into account the azimuthal acceptance also) of the produced near-projectile fragments 
falls in the angular acceptance  of the MARS spectrometer.  %%%as is also indicated quantitatively in figure 1. 
%%%
However, with an advanced large-acceptance separator like KOBRA, a larger fraction  of near-projectile 
fragments (that can reach 50--80\%) can be collected,  provided that we swing the primary beam at the 
appropriate angle (that we can choose to be near the  grazing angle of the reaction).

In figure 3(b), we show the mass resolved angular distributions for the reaction of 
$^{40}$Ar (15 MeV/nucleon) with the heavier target $^{238}$U. The grazing angle for this system 
is 20.0$^{o}$. We clearly see that the angular distribution of the near-projectile fragments
is much broader, making the efficient collection of the fragments especially challenging.
%%%
For this reaction,  the optimal angle to send the beam on the target (or, equivalently,
to rotate a large acceptance spectrometer) is  18$^{o}$, 
thus, collecting the fragments in the polar angular range of  13--23$^{o}$. 
%%%
With a solid angle acceptance of  $\Delta \Omega$ = 50 msr, the collection efficiency 
can reach 20--30\%.
%%%
We note, however, that the current design of the KOBRA spectrometer allows the beam to 
be sent on the target with a maximum angle of 12$^{o}$, in which case the collection 
efficiency may be lower by a factor of 2.
%%% 

From a practical standpoint, the use of the lighter target is preferable, as it leads to narrower 
angular distributions and thus larger collection efficiency, which may be desirable 
for certain applications.
%%%
However, the enhanced production of the most exotic nuclides obtained with the $^{238}$U target, 
encourages the development of approaches for the efficient collection of fragments 
from these reactions. 
%%%
%%% Of course a reasonable compromise between these two targets may be a $^{124}$Sn target.
%%%
In closing, we mention that our event-by-event simulations may allow full event tracking of the products
through the beam-optical elements of the separator.  Consequently rates for the production and separation 
of desired neutron-rich nuclides can be determined, as is currently being performed by members of 
the KOBRA team at RISP.
		
%%%----------------------------------------------------------------------------------	
%%%----------------------------------------------------------------------------------------------
%%%----------------------------------------------------------------------------------------------

\section{Reactions with a $^{48}$Ca  projectile at 15 MeV/nucleon }

The above calculations for the  $^{40}$Ar (15 MeV/nucleon) projectile are in fair agreement with
the experimental data giving us confidence on the reliability of the models used.
%%%
Moreover, the results of the reactions with $^{238}$U showed us that this target offers the possibility  
of enhanced production of the most neutron-rich nuclides.
%%%
Motivated by these promising results, we proceeded to perform calculations with the most neutron-rich stable beam 
in this mass range, namely $^{48}$Ca at 15 MeV/nucleon for which there are no experimental data at present. 
	
%%%--------------------------------------------------------------------------------------------
%%% 48Ca+64Ni, DIT/SMM,  CoMD/SMM  comparison.  Figure  remove
%%%
%%% In figure..., we show  the calculated mass distributions of projectile-like fragments with Z=14-21 from the 
%%% reaction %$^{48}$Ca (15 MeV/A) + $^{64}$Ni performed by DIT/SMM (solid red line), CoMD/SMM (blue dashed line) and 
%%%  CsMD/SMM (green dashed-dotted line).We observe that the three models lead to similar results, in the isotopes 
%%% near the %projectile, giving products that have captured up to 5 neutron. As before, we will proceed our 
%%% calculations with the DIT model. 
%%%--------------------------------------------------------------------------------------------
%%% 48Ca+ Ni, U   DIT/SMM, data 48Ca(140MeV/u)+Ta    figure 5
%%%
	
In figure 5, we show our DIT/SMM calculations for the reactions $^{48}$Ca (15 MeV/nucleon) + $^{64}$Ni 
[solid (red) line] and $^{48}$Ca (15 MeV/nucleon) + $^{238}$U [dashed (blue) line]. 
%%% 
In addition, we present the projectile fragmentation data of Mocko et al. \cite{Mocko-2006a,Mocko-2006b} for the
reaction $^{48}$Ca (140 MeV/nucleon) + $^{181}$Ta (open circles).

At first, observations similar to those for figure 3 pertain here.
%%%
We see that both the $^{48}$Ca+$^{64}$Ni and $^{48}$Ca+$^{238}$U reactions at 15 MeV/nucleon 
lead to substantial yields  of neutron pickup products with the latter reaction offering 
higher yields.
%%%
Interestingly, the fragmentation of $^{48}$Ca at 140 MeV/nucleon with the heavy neutron-rich 
$^{181}$Ta (N/Z=1.48) target leads to the production of neutron pickup products with cross sections
nearly similar to those calculated for the  $^{48}$Ca(15 MeV/nucleon)+$^{64}$Ni reaction. 
%%%
(We mention that such nuclides are also produced in the fragmentation of $^{48}$Ca (140 MeV/nucleon) 
on the  $^{9}$Be target with cross ections lower by at least a factor of 10.)
%%%
It would be interesting to describe theorerically the neutron pickup products from 
typical fragmentation reactions on heavy targets (e.g. $^{181}$Ta, $^{208}$Pb, $^{238}$U)  at the energy
of $\sim$100 MeV/nucleon,  which is at the lower limit of the typical projectile fragmentation mechanism 
\cite{Mocko-2008,EPAX3}. In the near future, we plan to undertake such a project using our theoretical 
model framework. 
%%%
In regards to higher energy projectile fragmentation, for completeness, we report the work of Suzuki et al.,
\cite{Suzuki-2013} on $^{48}$Ca (345 MeV/nucleon) + $^{9}$Be,  in which no pickup products were 
reported. The most neutron-rich isotopes observed were proton-removal products with cross sections
similar to those of the 140 MeV/nucleon data of Mocko et al. \cite{Mocko-2006a}. 

%%%**********************************************************************************************************************

\section{Reactions with the radioactive beams of  $^{46}$Ar and $^{54}$Ca at 15 MeV/nucleon}

After the above calculations and comparisons on reactions with the stable beams of $^{40}$Ar and  $^{48}$Ca,
we continue our study with reactions involving the neutron-rich radioactive beams (RIB) of $^{46}$Ar (Figs. 6 and 7)
and $^{54}$Ca (Figs. 8 and 9). We chose these neutron-rich projectiles as representative examples of 
RIB in this mass range, having six more neutrons than their stable counterparts  
$^{40}$Ar and  $^{48}$Ca, respectively.

%%%--------------------------------------------------------
%%% 40Ar, 46Ar + 64Ni DIT/SMM Fig. 6
%%%

In figure 6, we show the DIT/SMM calculations for the RIB reaction $^{46}$Ar (15 MeV/nucleon) + $^{64}$Ni 
[dashed (blue) line]  compared  with the calculation for the stable-beam reaction 
$^{40}$Ar (15 MeV/nucleon) + $^{64}$Ni and our   experimental data (closed points) as in figure 2.
%%%
We observe that the RIB leads to an isotope distribution with a neutron-rich side displaced by several 
neutrons (nearly six for near-projectile elements) compared with that of the stable beam. We also notice that 
the enhanced production of the neutron-rich isotopes offered by the RIB diminishes substantially for products
further away from the projectile (see also \cite{Fountas-2014}, figure 9).	
%%%--------------------------------------------------------
%%% 46Ar + 64Ni, U  DIT/SMM Fig. 7
%%%
In figure 7, we show the DIT/SMM calculations for the reactions of $^{46}$Ar (15 MeV/nucleon) with $^{64}$Ni 
[solid (red) line] and $^{238}$U [dashed (blue) line]. As expected,  we notice that the $^{238}$U target leads
to enhanced neutron-rich isotope production,  compared with the $^{64}$Ni target, especially at the tails 
of the distributions for near-projectile elements.

%%%--------------------------------------------------------
%%% 48Ca, 54Ca + 64Ni DIT/SMM figure 8
%%%
Continuing our discussion, in figure 8, we show the DIT/SMM calculations for the RIB reaction 
$^{54}$Ca (15 MeV/nucleon) + $^{64}$Ni [dashed (blue) line] compared with the calculations  for the 
stable-beam reaction $^{48}$Ca (15 MeV/nucleon) + $^{64}$Ni [solid (red) line]. As in figure 6, we observe
that the RIB results in an isotope distribution with a neutron-rich side displaced to the right compared 
with that of the stable beam.
%%%--------------------------------------------------------
%%% 54Ca + 64Ni, U   DIT/SMM figure  9
%%%
Finally, in figure 9, we compare the DIT/SMM calculations for the reactions of $^{54}$Ca (15 MeV/nucleon) with $^{64}$Ni 
[solid (red) line] and $^{238}$U [dashed (blue) line]. Observations similar to those of figure 7 pertain here. 
Moreover, we notice that extremely neutron-rich and even new isotopes toward $^{60}$Ca are produced
in the reaction with the $^{238}$U target.

%%%**********************************************************************************************	%%%**********************************************************************************************
%%% RIB rates:

\section{Isotope production rates}

A comprehensive presentation on the Z--N plane of the DIT/SMM calculated production cross sections
of  projectile  fragments from the 15 MeV/nucleon reactions  $^{48}$Ca + $^{238}$U and  
$^{54}$Ca + $^{238}$U is given in figure 10. 
%%%
In this figure, stable isotopes are represented by closed squares, whereas fragments obtained by
the respective reactions are given by the open circles (with sizes corresponding to cross-section 
ranges according to the figure key). 
%%%
The solid (red) line gives the expected location of the neutron drip line 
as calculated in Ref. \cite{halflife2}. 
%%%
In the figure, we clearly observe that the neutron pickup products from the radioactive beam 
reaction (figure 10b) extend toward the region of the neutron drip line near $^{62}$Ca.

We now proceed to a discussion of the production rates of some representative neutron-rich rare isotopes. 
%%%
In table 1,  we show the DIT/SMM calculated cross sections and the predicted production rates of some isotopes 
from the reaction $^{48}$Ca (15 MeV/nucleon) + $^{238}$U. 
%%%
We assume a beam intensity of 500 particle nA (3x10$^{12}$ particles/s) and a target thickness of 20 mg/cm$^{2}$.  

%%%**************************************************************************************************************
%%%**************************************************************************************************************	
%%% Table 1 **************************************************************************************	

\begin{table}[t]                     %%%%%%[tbph]
\caption{ Calculated cross sections (with DIT/SMM) and rates of neutron-rich isotopes
          from the stable-beam reaction $^{48}$Ca (15 MeV/nucleon) + $^{238}$U.
          For the rates, the beam intensity is assumed to be 500 pnA (3$\times$10$^{12}$ particles/sec)
          and the target thickness 20 mg/cm$^{2}$. Isotope halflives are taken from \cite{halflife1,halflife2}.
          }
%-------------------------
\vspace{0.5cm}
%-------------------------
\begin{center}
\begin{tabular}{ |c|c|c|c|c|} 
 \hline
 Rare isotope & t$_{1/2}$ (s) & Reaction channel & Cross section (mb) & Rates (s$^{-1}$)   \\
 \hline 
 $^{54}$Ca    & 0.09          & -0p + 6n         &  0.030              & 4.6 x 10$^{3}$  \\ 
%%\hline
 $^{46}$Ar    & 8.4           & -2p + 0n         &  2.9               & 4.4 x 10$^{5}$  \\ 
%%\hline
 $^{55}$Sc    & 0.09          & +1p + 6n         &  0.050              & 7.8 x 10$^{3}$  \\ 
%%\hline
 $^{52}$K     & 0.10          & -1p + 5n         &  0.050              & 7.8 x 10$^{3}$  \\
 \hline
%%%
\end{tabular}
\end{center}
\end{table}

%%%**************************************************************************************************************
%%%**************************************************************************************************************
%%***************************************************************************************************************

Subsequently, in table 2, we show the predicted  cross sections and production rates of several isotopes
from the reaction of a radioactive beam of $^{46}$Ar (15 MeV/nucleon) with $^{238}$U. In this case, 
we assumed that the beam intensity is equal to the production rate of  $^{46}$Ar from the reaction $^{48}$Ca 
(15 MeV/nucleon) + $^{238}$U  (table 1) and that the target thickness is again 20 mg/cm$^{2}$. 
We see that very exotic nuclei such as $^{52}$Ar and $^{49}$Cl can be produced with rates that 
may allow spectroscopic studies.

%%%**************************************************************************************************************
%%%**************************************************************************************************************	
%%%Table 2 -----------------------------------------------------------------------------	
%%%
\begin{table}[t]                     %%%%%%[tbph]
\caption{ Calculated cross sections (with DIT/SMM) and rates of neutron-rich isotopes
          from the radioactive-beam reaction $^{46}$Ar (15 MeV/nucleon) + $^{238}$U.
          For the rates, the beam intensity is assumed to be 4.4$\times$10$^{5}$ particles/sec
          and the target thickness 20 mg/cm$^{2}$.
          Isotope halflives are taken from \cite{halflife1,halflife2}.
          }
%-------------------------
\vspace{0.5cm}
%-------------------------
%%%%
\begin{center}
\begin{tabular}{ |c|c|c|c|c|} 
 \hline
 Rare isotope & t$_{1/2}$ (ms) & Reaction channel & Cross section (mb) & Rates (h$^{-1}$)   \\
 \hline 
 $^{51}$Ar    &   24           & -0p + 5n         &      0.064         &   5.0   \\ 
%%\hline
 $^{52}$Ar    &   16           & -0p + 6n         &      0.008         &   0.6   \\ 
%%\hline
 $^{48}$Cl    &   39           & -1p + 3n         &      0.24          &   18    \\ 
%%\hline
 $^{49}$Cl    &   28           & -1p + 4n         &      0.064         &   5.0   \\
 \hline
%%% 
\end{tabular}
\end{center}
\end{table}

%%%**************************************************************************************************************
%%%**************************************************************************************************************

Finally, in table 3, we show the predicted  cross sections and production rates of nuclides
from  the reaction of a radioactive beam of $^{54}$Ca (15 MeV/nucleon) with $^{238}$U. 
As before, we assumed a beam intensity  equal to the production rate of $^{54}$Ca from the 
reaction $^{48}$Ca (15 MeV/nucleon) + $^{238}$U (table 1) and a target thickness of 20 mg/cm$^{2}$.
%%%
We notice that relatively low but "usable" rates of extremely rare and even new isotopes,
like $^{59}$Ca ($\sim$1 per day) and $^{60}$Ca ($\sim$1 per week) can be produced that may allow 
to obtain identification and basic spectroscopic information for such nuclides.

%%%**************************************************************************************************************
%%%**************************************************************************************************************
%%%Table 3 --------------------------------------------------------------------------
%%%
\begin{table}[t]                     %%%%%%[tbph]
\caption{Calculated cross sections (with DIT/SMM) and rates of neutron-rich isotopes
          from the radioactive-beam reaction $^{54}$Ca (15 MeV/nucleon) + $^{238}$U.
          For the rates, the beam intensity is assumed to be 4.6$\times$10$^{3}$ particles/sec
          and the target thickness 20 mg/cm$^{2}$.
          Isotope halflives are taken from \cite{halflife1,halflife2}. (Theoretical estimates 
          for new nuclides are marked with an asterisk.)
          }
%-------------------------
\vspace{0.5cm}
%-------------------------
%%%%
\begin{center}
\begin{tabular}{ |c|c|c|c|c|} 
 \hline
 Rare isotope & t$_{1/2}$ (ms) & Reaction channel & Cross section (mb) & Rates (d$^{-1}$)   \\
 \hline 
 $^{57}$Ca    &  7             &  -0p + 3n        &   0.60             &   12       \\ 
%%\hline
 $^{58}$Ca    & 12             &  -0p + 4n        &   0.16             &   3.2       \\ 
%%\hline
 $^{59}$Ca    &  6$^*$         &  -0p + 5n        &   0.040            &   0.80      \\ 
%%\hline
 $^{60}$Ca    &  4$^*$         &  -0p + 6n        &  0.008             &   0.16      \\
%%\hline
 $^{54}$K     & 10             &  -1p + 1n        &   0.57             &    11       \\ 
%%\hline
 $^{55}$K     &  4             &  -1p + 2n        &   0.13             &   2.6      \\
 \hline
%%% 
\end{tabular}
\end{center}
\end{table}

%%%**************************************************************************************************************
%%%**************************************************************************************************************

\section{Discussion and Plans}

We would like to conclude the present study with some comments on the model approaches 
used in this work.
%%%
Starting from the microscopic CoMD model used in the dynamic stage, 
we think that the successful description of the reactions is especially 
valuable due to the predictive power of the microscopic many-body approach, 
as we have also seen in our recent works \cite{Fountas-2014,Vonta-2015,Vonta-2016}, 
that does not depend  on ad hoc assumptions of the reaction dynamics.
%%%
%%% However, the full CoMD description  is very computer intensive  and not practical, 
%%% but, essentially validates the two-step approach as we saw in this work.
%%%
We saw that the two-stage CoMD/SMM approach provided  good results in comparison
with the experimental data. Nonetheless, we point out that it is very computer intensive 
due to the N-body CoMD stage of the calculation \cite{Fountas-2014}, that limited its
extensive application  in this work.
%%%
On the  contrary, the phenomenological DIT/SMM approach, offering results
similar to the microscopic CoMD/SMM approach, is rather fast. For this reason we 
used it extensively in this study to estimate cross sections of very neutron-rich
products. Moreover, this approach can be practical for the  design of RIB experiments
based on the present reactions.

%%% Continuation:
%%%
As a continuation of the present work, we plan to perform systematic CoMD calculations
toward  the region of very low cross sections and compare them with the corresponding DIT
results. Furthermore, we plan to continue the systematic calculation efforts with 
radioactive beams that are  expected from upcoming RIB facilities in the energy 
range of 10--25 MeV/nucleon.
%%%
Concerning the choice of energy, our calculations,  presented in the appendix, suggest that
the energy of 15 MeV/nucleon is a reasonable choice for the efficient production of 
very neutron-rich isotopes close to the projectile.
%%%
This conclusion is also in agreement with our pervious experimental \cite{GS-PRC-2011}
and theoretical \cite{Fountas-2014} work. 
%%%
Since our calculations are complete event-by-event simulations, we can also 
study systematically the velocity distributions, the angular distributions and,
the ionic charge state distributions of the projectile fragments.
%%%
This information can form the basis of realistic beam optics simulations of the 
behavior of large acceptance separators   and may help us optimize our ability  
to separate  and identify exotic neutron-rich nuclides. 

In the near future, apart from systematic RIB calculations  in the mass range 40--60,
we plan to perform detailed measurements of projectile fragments from $^{48}$Ca
and $^{70}$Zn beams at 10--15 MeV/nucleon  a) at Texas A\&M with the MARS recoil
separator  \cite{GS-NIM-2008,GS-PRC-2011}  and b) at LNS/INFN  with the MAGNEX large-acceptance 
spectrometer \cite{MAGNEX}.
%%%
We expect that these measurements will provide  a further detailed testing ground  for our models, 
and will offer access to very neutron-rich nuclei for a broad range of studies.
%%%
These studies may include measurements of secondary reactions, in-beam 
spectroscopy of selected products, decay measurements and further search for new isotopes
or new isomers. 
%%%
In parallel, these efforts will provide experience and preparation for future plans
at upcoming large-acceptance separator facilities
(e.g. KOBRA \cite{KOBRA-2016}, ISLA \cite{ISLA}).
%%%

%%%************************************************************************
%%%************************************************************************

\section{Summary and conclusions}  
%%%************************************************************************

In this article we report on our continued efforts to systematically study 
the production of neuton-rich rare isotopes with heavy-ion beams in the energy range
of 15--25 MeV/nucleon. 
%%%----------------
We studied the production of neutron-rich projectile-like isotopes in multinucleon
transfer collisions of stable and radioactive beams in the mass range A$\sim$40--60. 
%%%
We first presented our experimental cross section data on $^{40}$Ar(15 MeV/nucleon) + $^{64}$Ni,
$^{58}$Ni and $^{27}$Al and compared them with calculations.
%%%
The calculations are based on two models: the phenomenological  deep-inelastic transfer (DIT)
model and the microscopic constrained molecular  dynamics (CoMD) model  employed to describe
the dynamical stage of the collision.  De-excitation of the resulting projectile-like
fragments is performed with the statistical multifragmentation model (SMM). 
An overall good agreement of the calculations with the experimental data is obtained.  
%%%
We also performed calculations of the reaction of $^{40}$Ar (15 MeV/nucleon) projectile
with the heavy neutron-rich target of $^{238}$U. We then continued with the 
reactions of the neutron-rich beam of $^{48}$Ca (15 MeV/nucleon)  with targets of
$^{64}$Ni and $^{238}$U. 
%%%
In these reactions,  we observed that neutron-rich rare isotopes with sustantial cross sections can
be produced which, in turn, may be assumed to form radioactive beams and interact with a subsequent target, 
preferably $^{238}$U, resulting in extremely neutron-rich and even new isotopes 
(e.g. $^{60}$Ca) in this  mass range.   %%%% of A=40--60.
%%%
We conclude that multinucleon transfer reactions with stable or radioactive beams 
at the energy of around 15 MeV/nucleon  constitute a novel and competitive route to access extremely
neutron-rich rare isotopes  that may open up exciting  nuclear research opportunities
in current or upcoming rare isotope beam facilities.
%%%

%%**************************************************************************************
%%**************************************************************************************
%%**************************************************************************************

\section{Acknowledgements}

%%%We gratefully acknowledge the support of the operations staff
%%%of the Cyclotron Institute during the measurements.
%%%.
We are thankful to L. Tassan-Got for the DIT code, M. Papa and  
for his version of the CoMD code and A. Botvina for the SMM code.
%%%
%% We thank A.L. Keksis, Z. Kohley and B.C. Stein
%% for the help and support at various stages of the experimental work 
%% on rare isotope production at 15-25 MeV/nucleon at the Cyclotron Institute of Texas A\&M 
%% University.
%%%
We also thank  A. Pakou, O. Sgouros and V. Soukeras for numerous enlighting  
discussions and suggestions.
%%%
Financial support for this work was provided, in part, by
the National and Kapodistrian University of Athens
under ELKE Research Account No 70/4/11395 
%%%
and, in part, by the US Department of
Energy through Grant No. DEFG02-93ER40773 and the Robert
A. Welch Foundation through Grant No. A-1266.
%%%
K.T., S.C.J and B.H.K and Y.K.K  were supported
by the Rare Isotope Science Project of the Institute for Basic Science
funded by the Ministry of Science, ICT and Future Planning 
and National Research Foundation of Korea.
%%%
M.V. was  supported by the Slovak Scientific Grant Agency under contracts 2/0105/11
and 2/0121/14 and by the Slovak Research and Development Agency under contract
APVV-0177-11.

%%%*********************************************************************************
%%%**********************************************************************************************
%%%  48Ca+ U  (DIT/SMM)   figure 12   10, 15, 25 MeV/nucleon
%%% Appendix A:

\section{APPENDIX: Energy dependence of production cross sections}

In this appendix, we examine the dependence of the cross sections of the neutron-rich
products on the projectile energy.
%%%
We consider the stable-beam reaction of $^{48}$Ca with the $^{238}$U target and 
performed DIT/SMM calculations at beam energies of 10, 15 and 25 MeV/nucleon. 
The results are presented in  figure 10 by dashed (blue), solid (black) and dotted (red) lines, respectively.
%%%
We observe that on the neutron-rich side, the reactions at the lower two energies, 
10 and 15 MeV/nucleon,   lead to nearly similar cross sections which are higher 
than those of the reaction  at 25 MeV/nucleon.
%%%
On the neutron-deficient side, the distributions of the 15 and 25 MeV/nucleon reactions 
are similar, indicating a possible similarity of the excitation energies of the
quasiprojectiles leading to these isotopes. At 10 MeV/nucleon, the neutron-deficient 
nuclide cross sections are lower as a result of lower excitation energies reached in 
this reaction compared to the higher energy ones.
%%%
The above comparisons lead us to the conclusion that the beam energy of 15 MeV/nucleon 
is  a reasonable choice for the efficient production of the most neutron-rich nuclides 
in these multinucleon transfer reactions. %%% of neutron-rich beams with the $^{238}$U target.

%%%***********************************************************************************
%%% End of Appendices
%%%	
%%%-----------------------------------------------------------------------------------------
%%%\section{REFERENCES}

%***************************************************************************
%***************************************************************************
% All figures here:               Initially they were with  width=0.65 
%***************************************************************************
%***************************************************************************
%***************************************************************************

%%% Fig 1     Exp data 40Ar+64Ni,58Ni,27Al

\begin{figure}[hp]   %%% htbp
\begin{center}  
\includegraphics[width=0.85\textwidth,keepaspectratio=true]{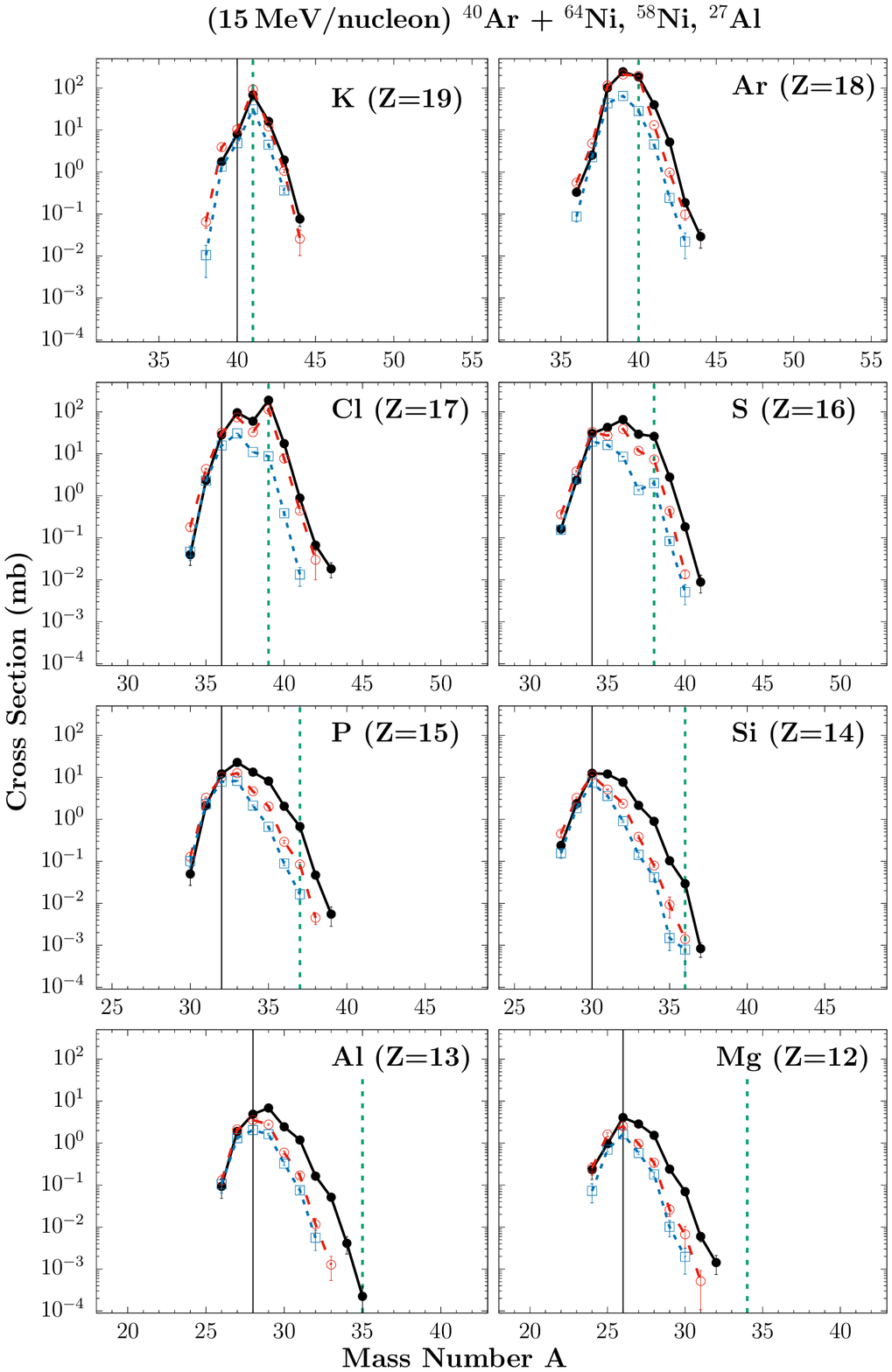}
\end{center}  
\caption{ (Color online)
%%%
Experimental mass distributions (cross sections) of the isotopes of elements Z=19--12 for 
the three reactions measured: $^{40}$Ar (15 MeV/nucleon) + $^{64}$Ni, $^{58}$Ni and $^{27}$Al,  
represented by closed (black) circles, open (red) circles and open (blue) squares, respectively.
%%%
Nuclides to the left of the thin solid lines are not fully covered by the magnetic
rigidity range of the experiment (see text).
%%%
Nuclides with a net pickup of neutrons from the target lie to the right of the 
dotted lines.   
%%%
}
\label{figure01}
\end{figure}

%%**********************************************************************************************
%%**********************************************************************************************

%Fig. 2   40Ar+Ni   DIT/SMM, CoMD/SMM  

\begin{figure}[htbp]                                        %%% htbp
\begin{center}  
\includegraphics[width=0.85\textwidth,keepaspectratio=true]{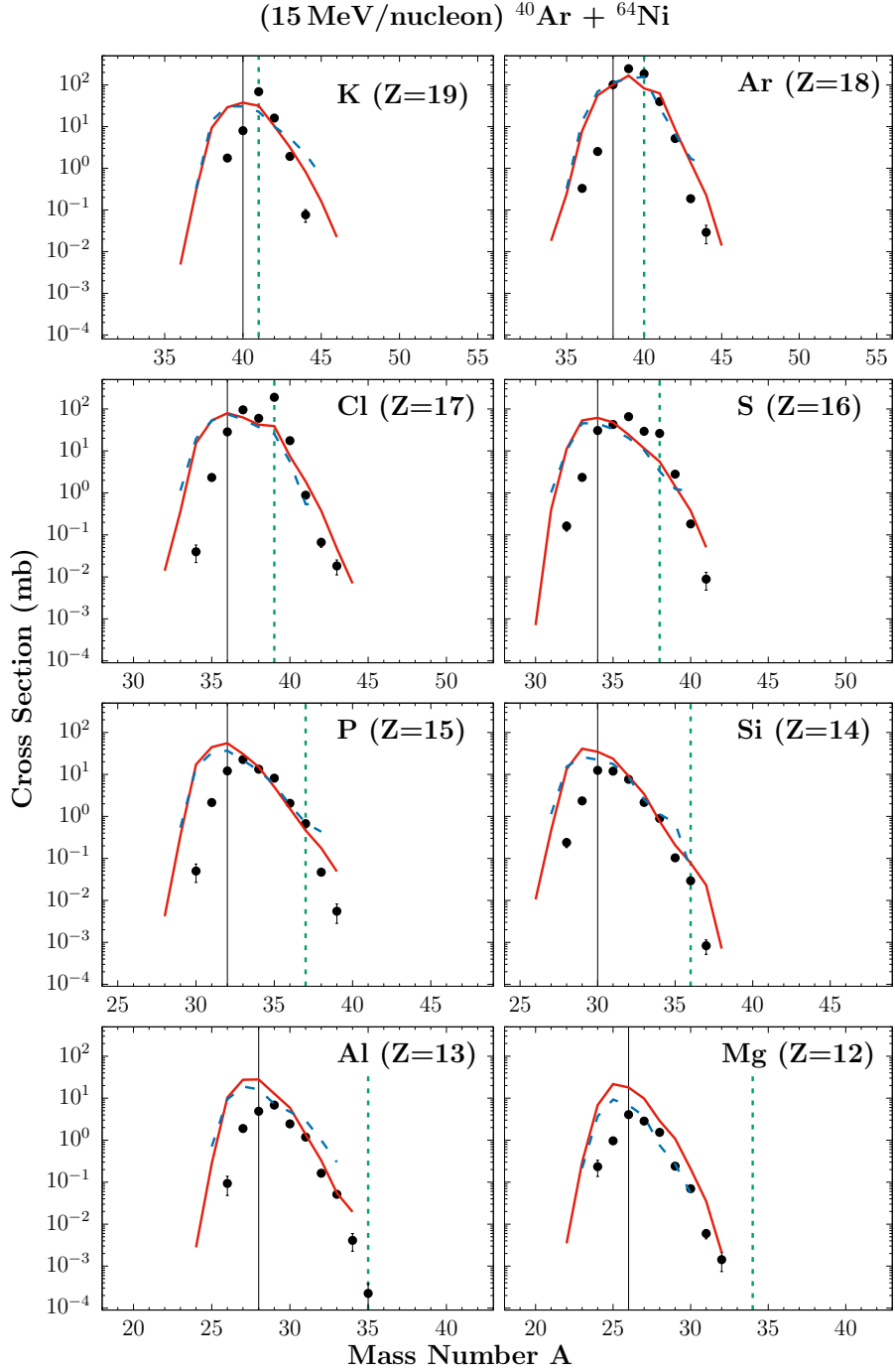}
\end{center}  
\caption{ (Color online)
%%%
Comparison of calculated mass distributions (lines) of projectile fragments with Z=19--12
from the reaction $^{40}$Ar (15 MeV/nucleon) + $^{64}$Ni with the experimental data 
[closed (black) circles, as in figure 1]. 
The calculations are: DIT/SMM [solid (red) line] and CoMD/SMM [dashed (blue) line].
%%%
Nuclides to the left of the thin solid lines are not fully covered by the magnetic
rigidity range of the experiment (see text).
%%%
Nuclides with a net pickup of neutrons from the target lie to the right of the 
dotted lines.
}
\label{figure02}
\end{figure}

%%*********************************************************************************************
%**********************************************************************************************

%Fig. 3    40Ar + Ni, Pb, U

\begin{figure}[htbp]                                        %%% htbp
\begin{center}  
\includegraphics[width=0.85\textwidth,keepaspectratio=true]{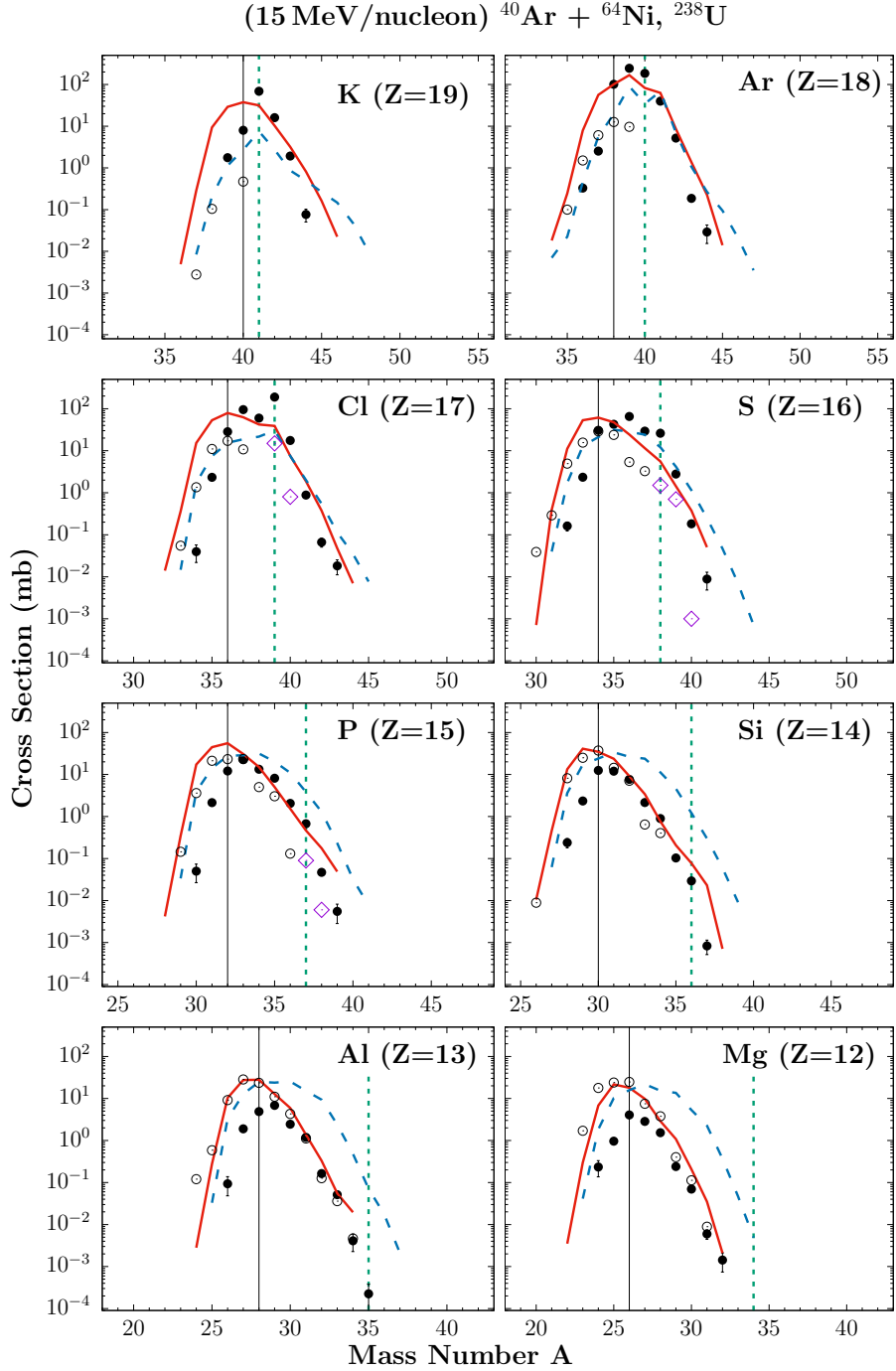}
\end{center}  
\caption{ (Color online)
%%%
DIT/SMM calculated mass distributions (cross sections) of projectile fragments with Z=19--12 from  the reaction $^{40}$Ar
(15 MeV/nucleon) + $^{238}$U [dashed (blue) line] compared with those from $^{40}$Ar (15 MeV/nucleon) + $^{64}$Ni  
[solid (red) line] and the experimental data [closed (black) circles].  
%%%
Nuclides to the left of the thin solid lines are not fully covered by the magnetic
rigidity range of the experiment (see text).
%%%
Nuclides with a net pickup of neutrons from the target lie to the right of the
dotted lines.
%%%
In addition, the projectile fragmentation 
data of \cite{Notani-2007} on $^{40}$Ar (94 MeV/nucleon) + $^{181}$Ta  are presented by open circles
and those of \cite{Zhang-2012} of the same system at 57 MeV/nucleon by open diamonds.
}
\label{figure03}
\end{figure}

%****************************************************************************************

%Fig. 4  Ang. Dist:  40Ar + Ni, U 

\begin{figure}[htbp]                                        %%% htbp
\begin{center}  
\includegraphics[width=0.75\textwidth,keepaspectratio=true]{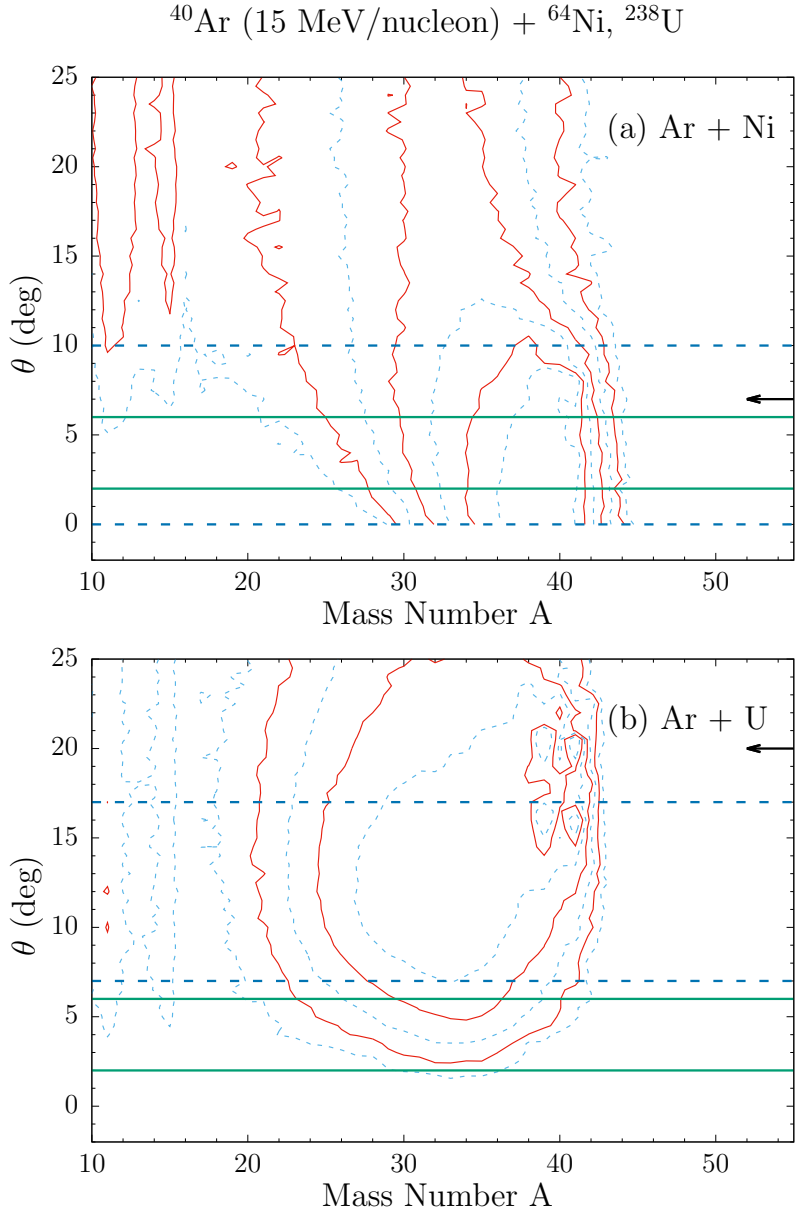}
\end{center}  
\caption{ (Color online)
%%%
(a) DIT/SMM calculated mass-resolved angular distibutions of projectile fragments 
from the reaction $^{40}$Ar (15 MeV/nucleon) + $^{64}$Ni.
The successive contours (starting from the innermost) represent a decrease in the yield
by a factor of two.
The  solid (green) lines represent the polar angular acceptance of the
MARS spectrometer, and  the dashed (blue) lines the optimum angular acceptance of the KOBRA
spectrometer for this reaction. The arrow indicates
the grazing angle of the Ar+Ni reaction (in the lab system).
%%%
(b) As in panel (a), but for the reaction  $^{40}$Ar (15 MeV/nucleon) + $^{238}$U.
In this case, the angular acceptance of KOBRA corresponds to the maximum angle
of the beam on the target (see text).
%%%
}
\label{figure04}
\end{figure}

%%%********************************************************************************************
%%%********************************************************************************************
%%%
%%% Fig. 5  48Ca + Ni, Pb, U    DIT/SMM,  48Ca(140MeV/u)+Ta data

\begin{figure}[htbp]                                        %%% htbp
\begin{center}  
\includegraphics[width=0.85\textwidth,keepaspectratio=true]{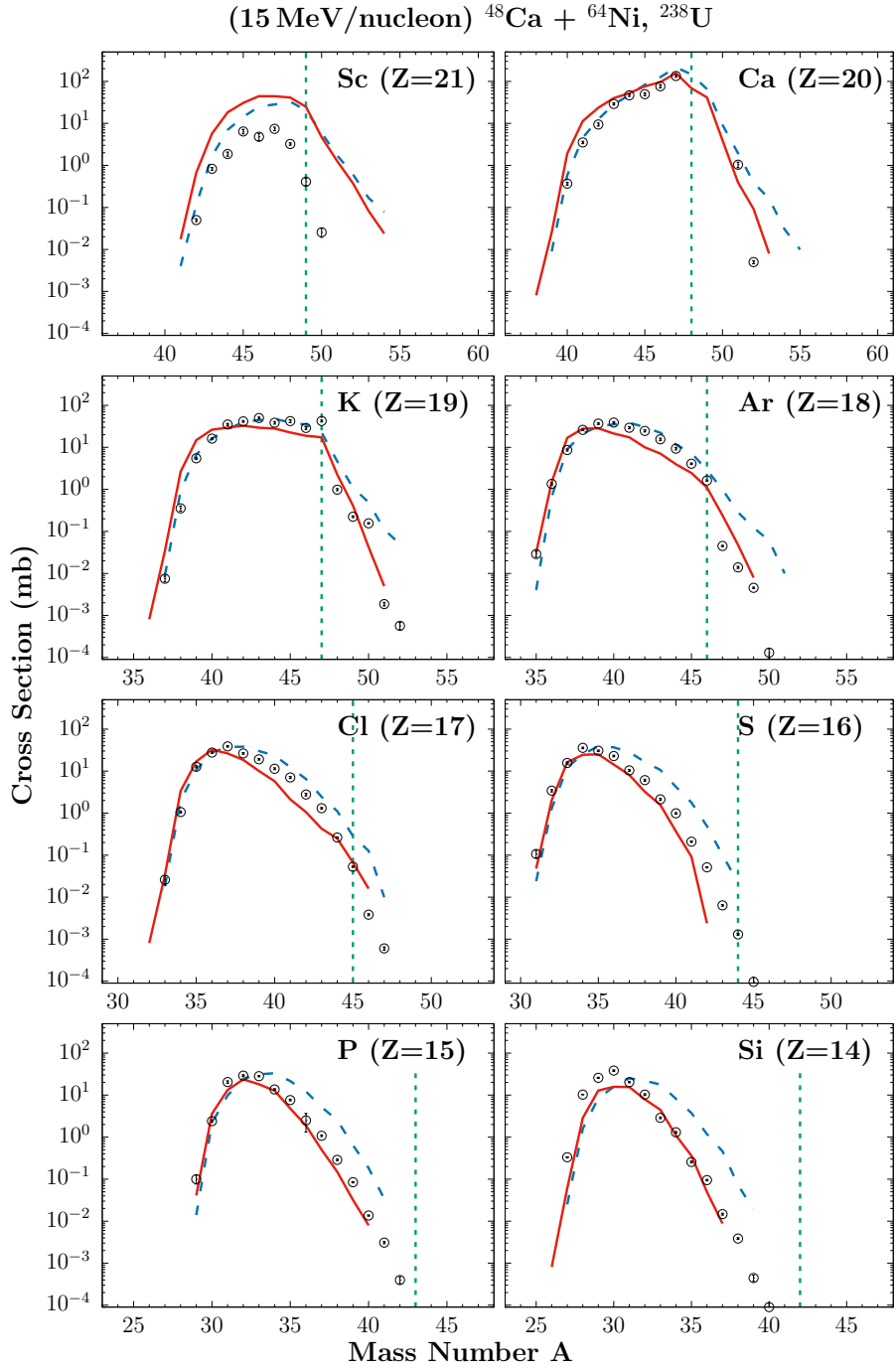}
\end{center}  
\caption{ (Color online)
%%%
DIT/SMM calculated mass distributions (cross sections) of projectile fragments with Z=21--14 
from  the reactions 
%%%
$^{48}$Ca (15 MeV/nucleon) + $^{64}$Ni [solid (red) line] and
$^{48}$Ca (15 MeV/nucleon) + $^{238}$U [dashed (blue) line]
%%%
compared with the projectile fragmentation data of Ref. \cite{Mocko-2006a,Mocko-2006b} on 
$^{48}$Ca (140 MeV/nucleon) + $^{181}$Ta [open (black) circles].
%%%
Nuclides with a net pickup of neutrons from the target lie to the right of the
dotted lines.
%%%
}
\label{figure05}
\end{figure}

%%%********************************************************************************************
%%%********************************************************************************************
%%%                RIB:
%%% Fig. 6   40Ar, 46Ar + Ni

\begin{figure}[htbp]                                        %%% htbp
\begin{center}  
\includegraphics[width=0.85\textwidth,keepaspectratio=true]{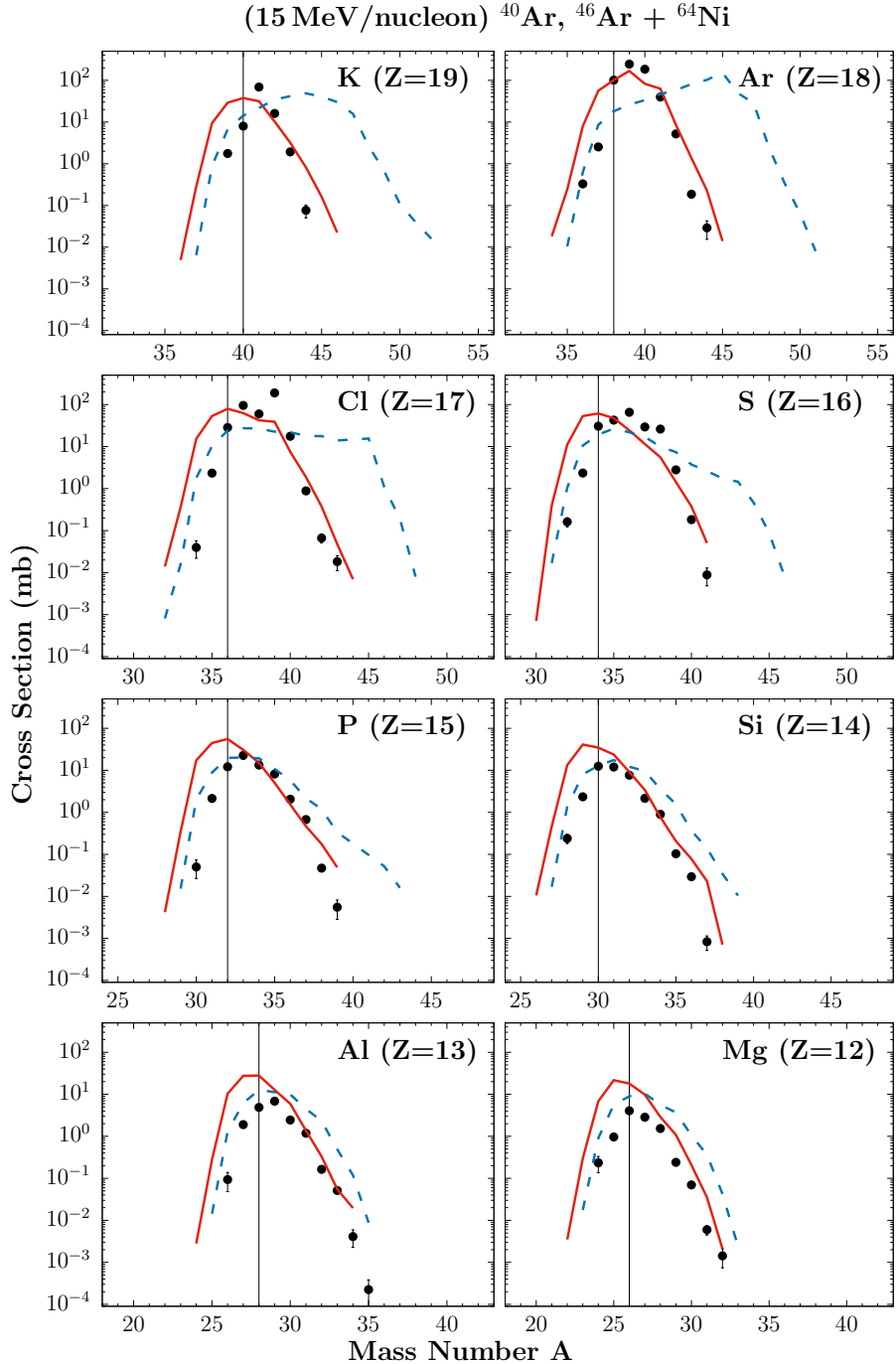}
\end{center}  
\caption{ (Color online)
%%%
DIT/SMM calculated mass distributions (cross sections) of projectile fragments with Z=19--12 
from the radioactive-beam reaction
$^{46}$Ar (15 MeV/nucleon) + $^{64}$Ni [dashed (blue) line] 
compared with the stable-beam reaction
$^{40}$Ar (15 MeV/nucleon) + $^{64}$Ni [solid (red) line]
and the experimental data [closed (black) circles].
%%%
As in Figs. 1--3, nuclides to the left of the thin solid lines are not fully covered by the magnetic
rigidity range of our 15 MeV/nucleon experiment.
%%%
%%%
}
\label{figure06}
\end{figure}

%%%********************************************************************************************
%%%********************************************************************************************
%%%
%%% Fig. 7   46Ar + Ni, U

\begin{figure}[htbp]                                        %%% htbp
\begin{center}  
\includegraphics[width=0.85\textwidth,keepaspectratio=true]{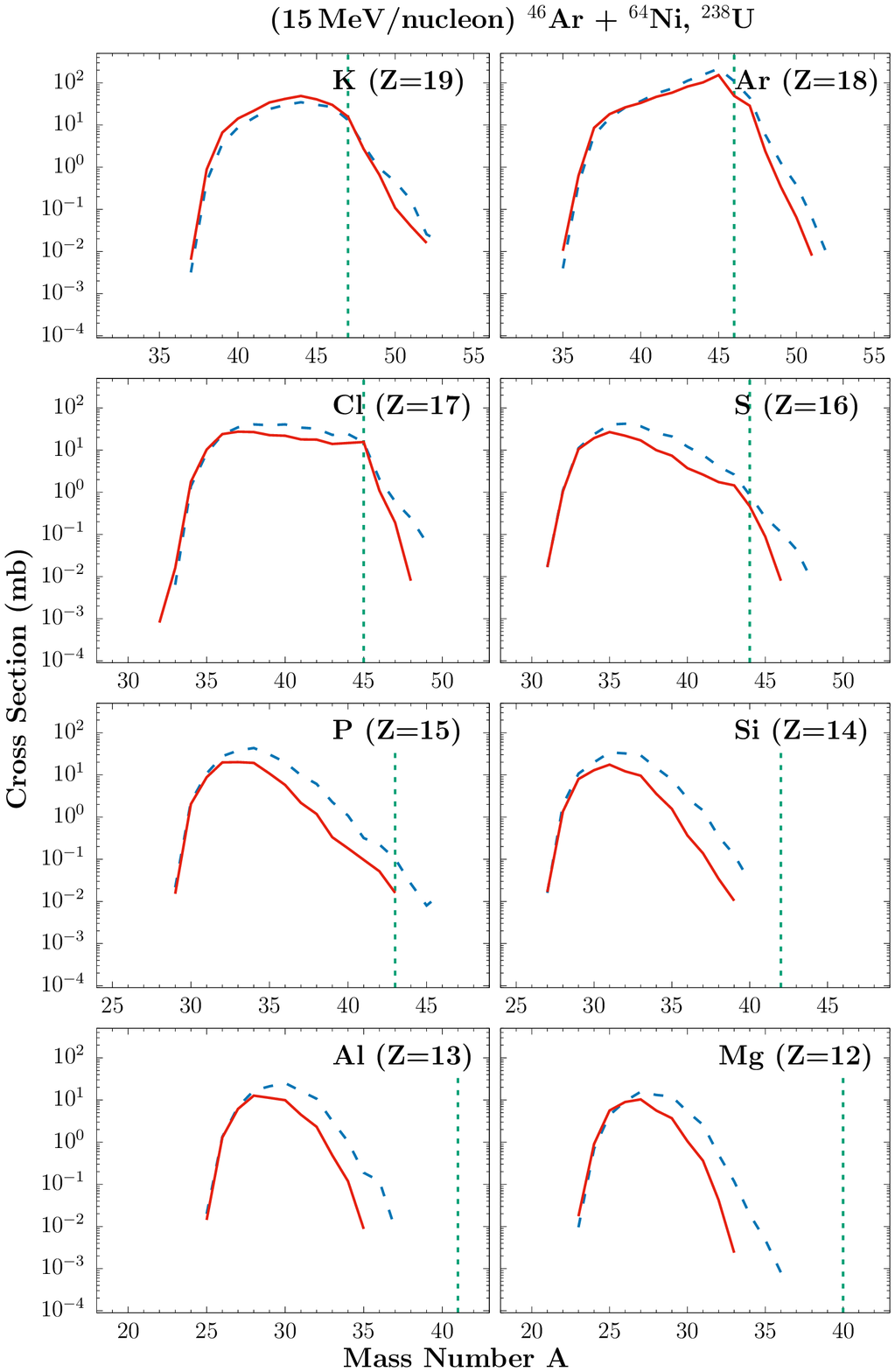}
\end{center}  
\caption{ (Color online)
%%%
DIT/SMM calculated mass distributions (cross sections) of projectile fragments with Z=19--12 
from the reaction of a radioactive-beam of $^{46}$Ar (15 MeV/nucleon) 
on $^{64}$Ni [solid (red) line] and $^{238}$U [dashed (blue) line].
Nuclides with a net pickup of neutrons from the target lie to the right of the
dotted lines.
%%%
}
\label{figure07}
\end{figure}

%%%********************************************************************************************
%%%********************************************************************************************
%%%
%%% Fig. 8    48Ca, 54Ca + Ni

\begin{figure}[htbp]                                        %%% htbp
\begin{center}  
\includegraphics[width=0.85\textwidth,keepaspectratio=true]{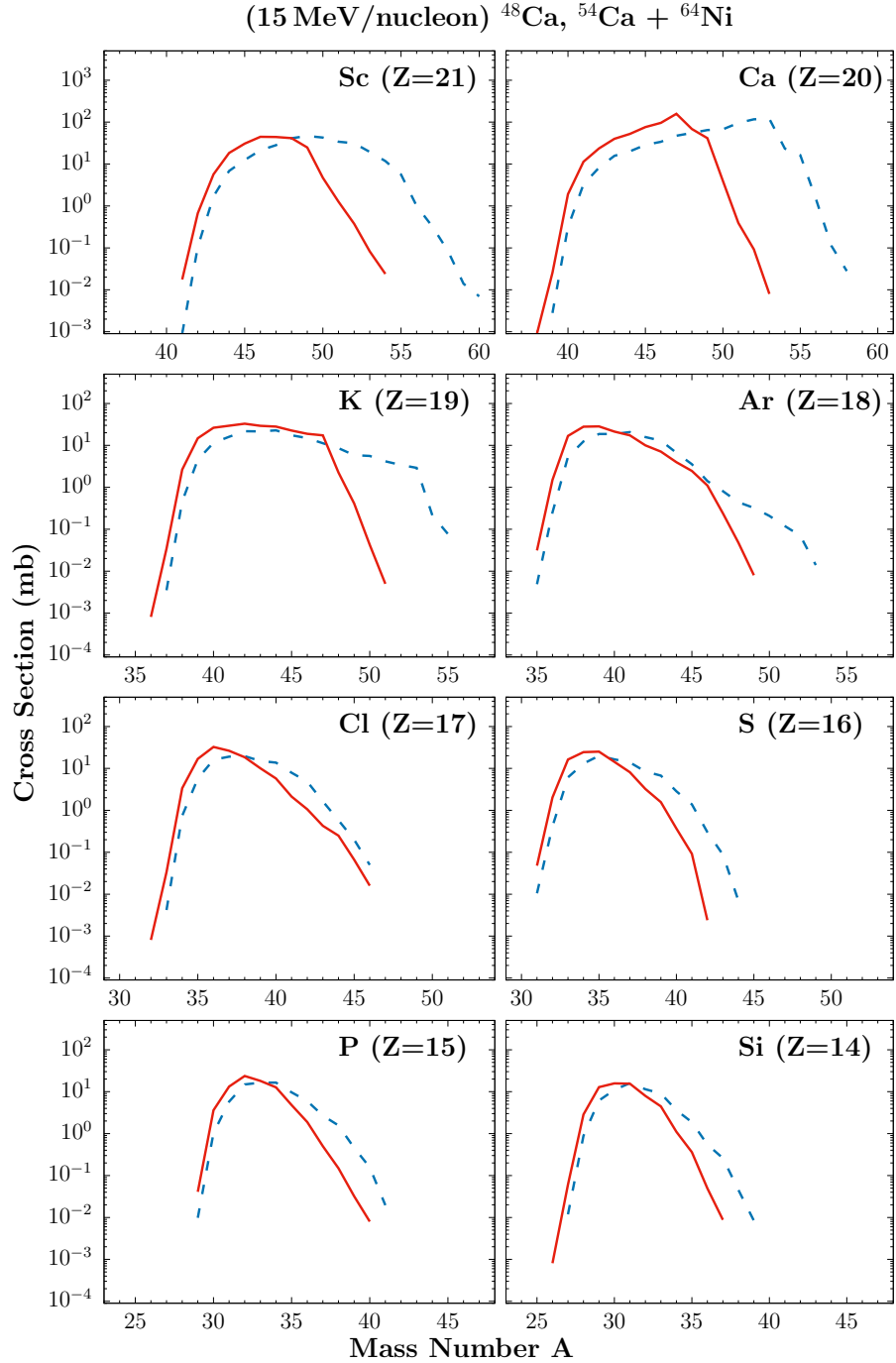}
\end{center}  
\caption{ (Color online)
%%%
DIT/SMM calculated mass distributions (cross sections) of projectile fragments with Z=21--14 
from the radioactive-beam reaction
$^{54}$Ca (15 MeV/nucleon) + $^{64}$Ni [dashed (blue) line] 
compared with the stable-beam reaction
$^{48}$Ca (15 MeV/nucleon) + $^{64}$Ni [solid (red) line].
%%%
}
\label{figure08}
\end{figure}

%%%********************************************************************************************
%%%********************************************************************************************
%%%
%%% Fig. 9   54Ca + Ni,U

\begin{figure}[htbp]                                        %%% htbp
\begin{center}  
\includegraphics[width=0.85\textwidth,keepaspectratio=true]{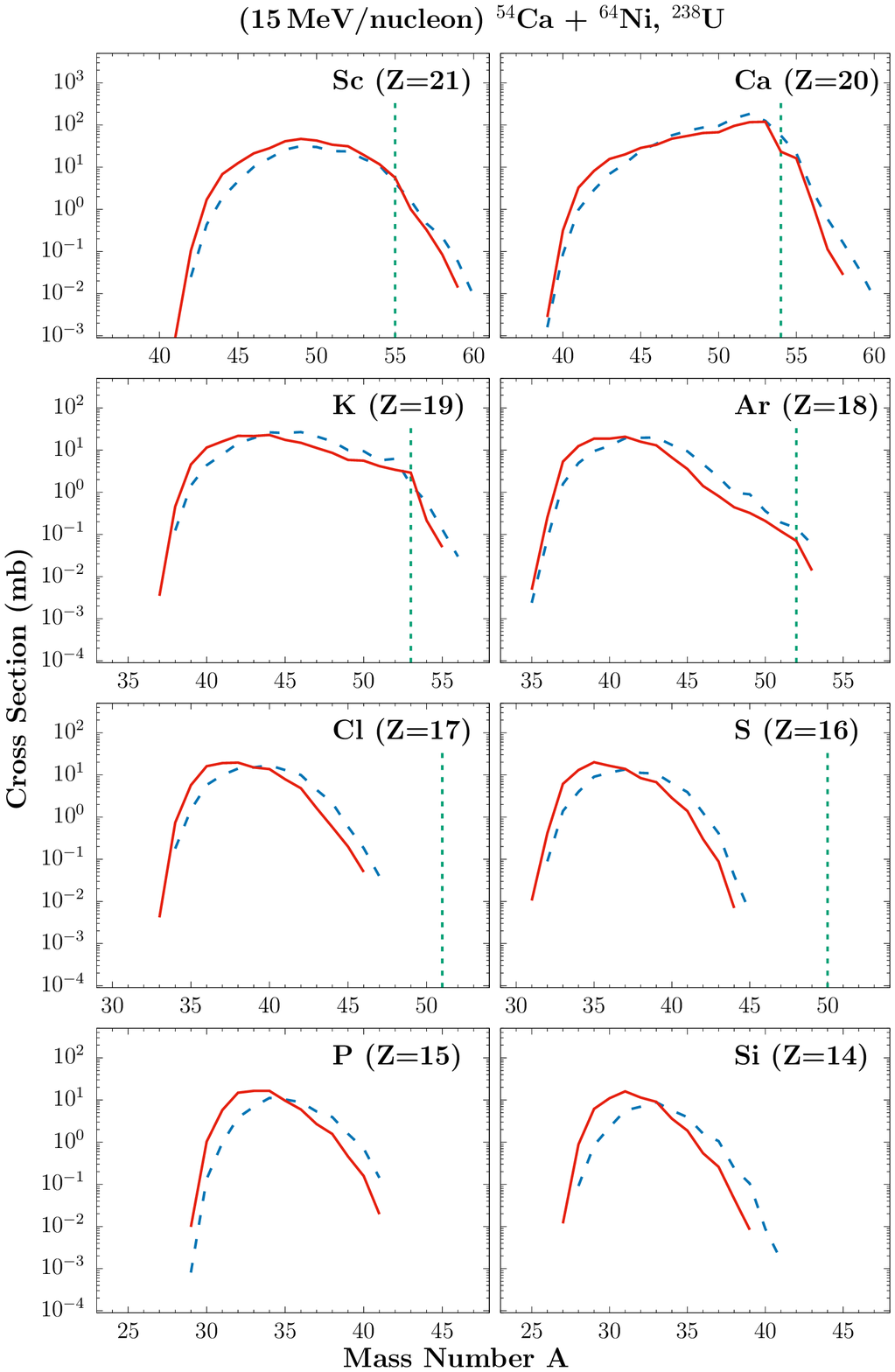}
\end{center}  
\caption{ (Color online)
%%%
DIT/SMM calculated mass distributions (cross sections) of projectile fragments with Z=21--14
from the reaction of a radioactive-beam of $^{54}$Ca (15 MeV/nucleon) 
on $^{64}$Ni [solid (red) line] and $^{238}$U [dashed (blue) line].
Nuclides with a net pickup of neutrons from the target lie to the right of the
dotted lines.
%%%
}
\label{figure09}
\end{figure}

%%********************************************************************************************
%*********************************************************************************************
%%%********************************************************************************************
%%%********************************************************************************************
%%% New: 
%%% Fig. 10   48Ca+U,  54Ca+U    Z-N plots

\begin{figure}[htbp]                                        %%% htbp
\begin{center}  
\includegraphics[width=0.85\textwidth,keepaspectratio=true]{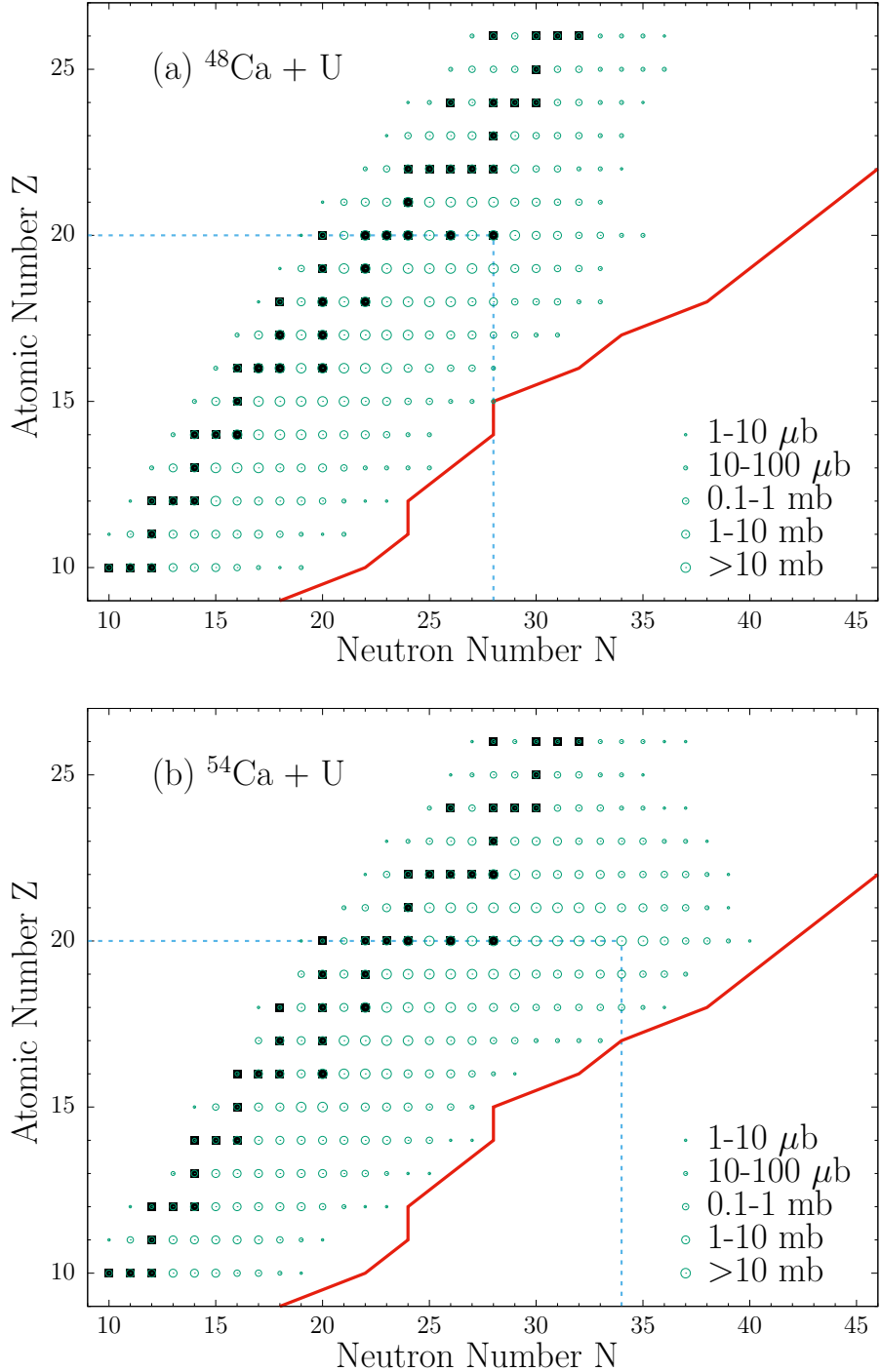}
\end{center}  
\caption{ (Color online)
%%%
(Color online) 
Representation on the Z--N plane of DIT/SMM calculated production cross sections of 
projectile fragments from 15 MeV/nucleon reactions of (a) the stable beam $^{48}$Ca 
with $^{238}$U, and (b) the radioactive-beam $^{54}$Ca with $^{238}$U. 
The cross section ranges are shown by open circles according to the key. 
The closed squares show the stable isotopes. The solid (red)
line represents the expected location of the neutron drip line according to \cite{halflife2}.    %%% Moller-1997 
The horizontal and vertical dotted line segments indicate the location of the 
projectiles.
%%%
}
\label{figure10}
\end{figure}

%%********************************************************************************************
%*********************************************************************************************
%%% Figures for the Appendix:

%%%********************************************************************************************
%%% Energy dependence:
%%%
%%% Fig. 11 (A1)  48Ca + U    10, 15, 25 MeV/nucleon

\begin{figure}[htbp]                                        %%% htbp
\begin{center}  
\includegraphics[width=0.85\textwidth,keepaspectratio=true]{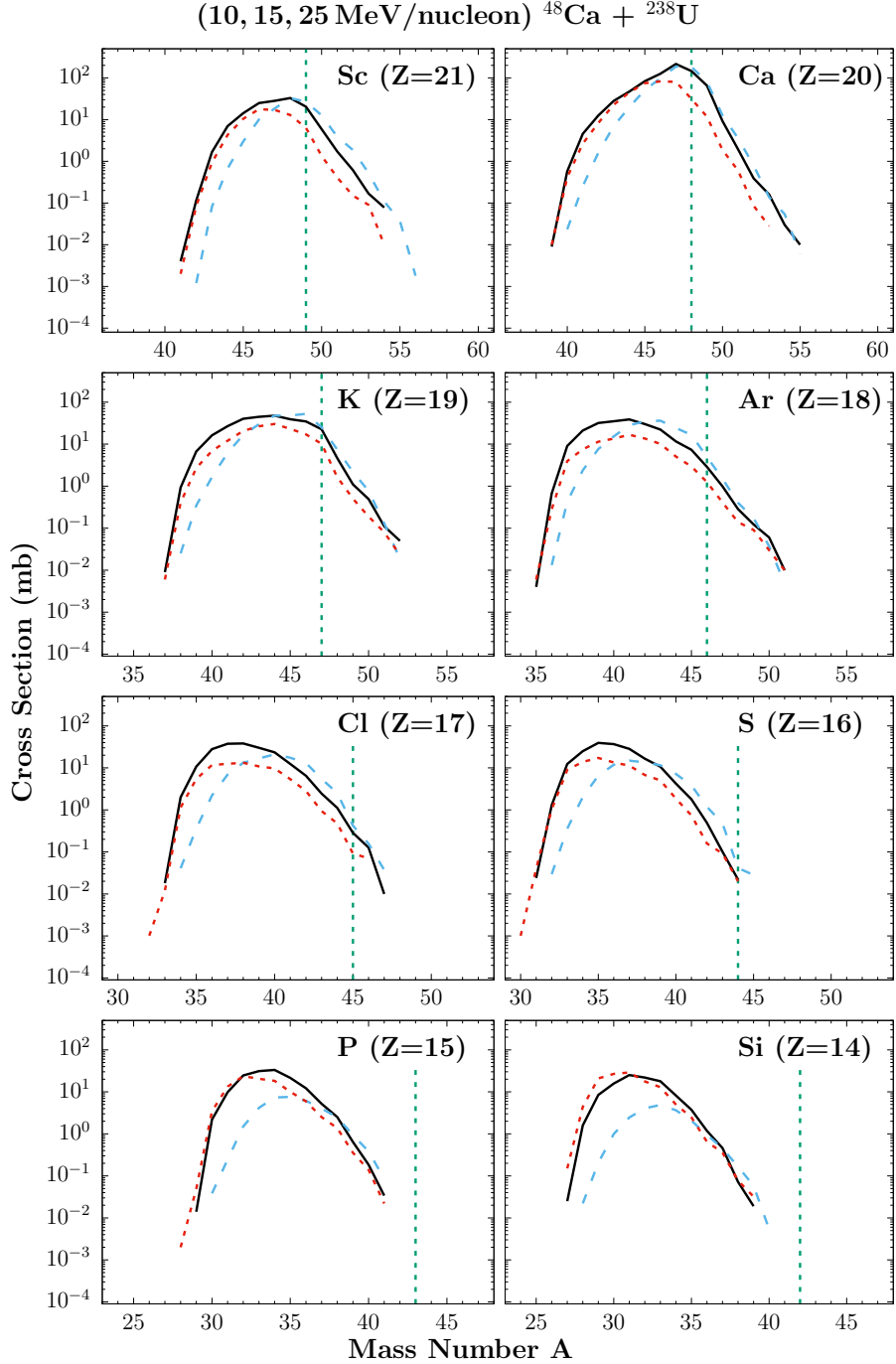}
\end{center}  
\caption{ (Color online)
%%%
Energy dependence of DIT/SMM calculated mass distributions of projectile 
fragments with Z=21--14 from the reaction of a $^{48}$Ca beam with a $^{238}$U target 
at energies of 10, 15 and 25 MeV/nucleon represented by dashed (blue), solid (black) and
dotted (red) lines respectively.
Nuclides with a net pickup of neutrons from the target lie to the right of the 
dotted lines.
%%%
}
\label{figure_a1}
\end{figure}

%%-------------------------------------------------------------------------------------
%%%********************************************************************************************
%%%********************************************************************************************

\end{document}